\documentclass[twocolumn,showpacs,preprintnumbers,amsmath,amssymb]{revtex4}

\usepackage{graphicx}
\usepackage{dcolumn}
\usepackage{bm}

\begin{document}

\preprint{APS/123-QED}

\title{
The hidden geometry of ocean flows}

\author{Carolina Mendoza}
\author{Ana M Mancho\cite{corr}}
\affiliation{%
Instituto de Ciencias Matem\'aticas, CSIC-UAM-UC3M-UCM, Serrano 121, 28006 Madrid, Spain.}%

\date{\today}

\begin{abstract}
We introduce a new global Lagrangian descriptor that is applied to flows with  general time dependence (altimetric datasets). It 
succeeds in  detecting  simultaneously, with great accuracy,  invariant manifolds, hyperbolic and non-hyperbolic flow regions.
\end{abstract}


\maketitle
{\it Introduction}.---   Finding order in the apparent disorder of ocean motion is still an open problem,  
  \begin{figure*}
\includegraphics[width=14.9cm]{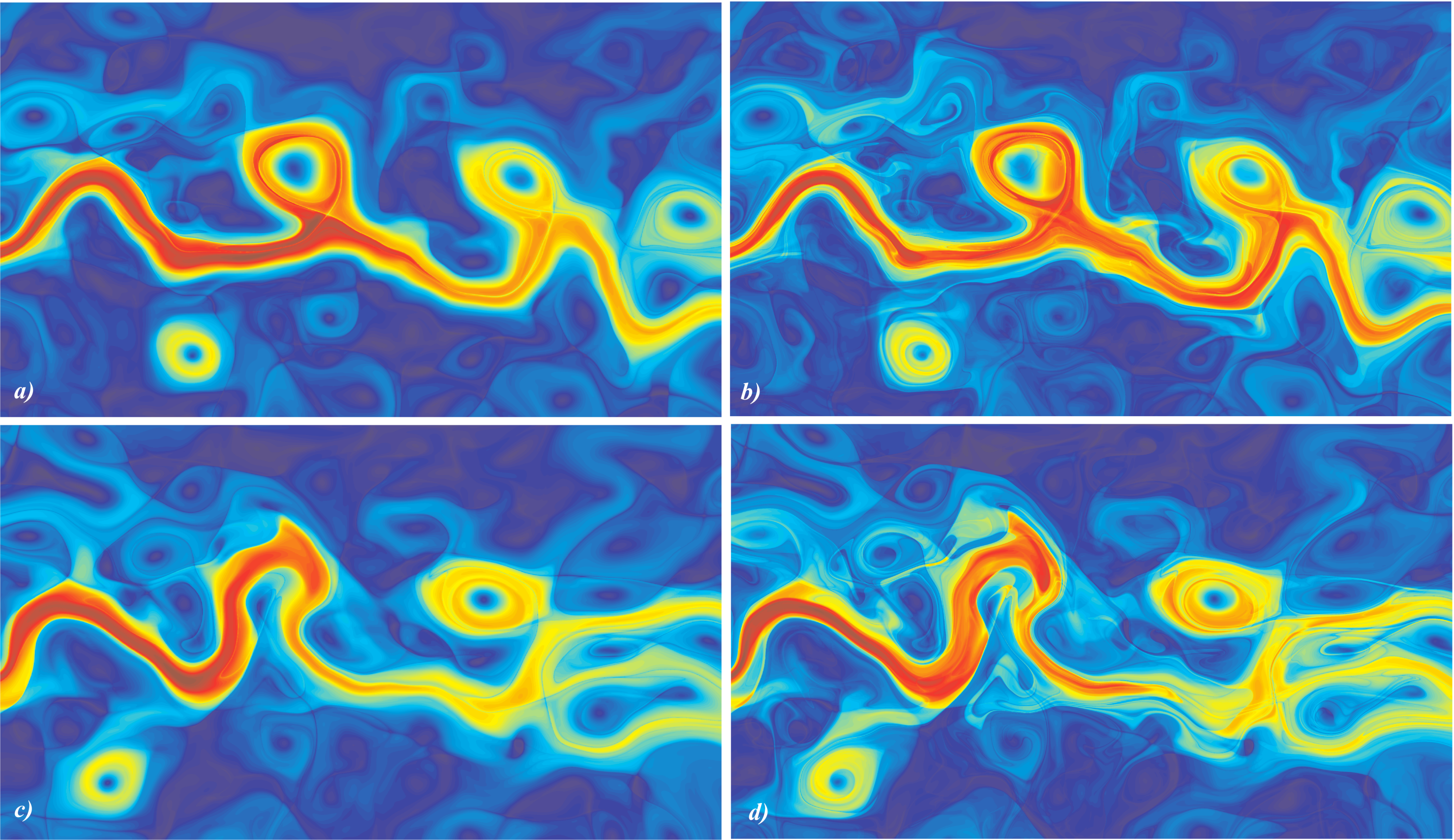}
\caption{\label{fig:M1}
Evaluation of the function $M$ over the Kuroshio current  between longitudes $148^o$E-$168^o$E and latitudes
 $30^o$N-$41.5^o$N;  a) and b) on  May 2, 2003; c) and d) on June 3, 2003. Panels a) and c) take $\tau=15$;
Panels b) and d) take $\tau=30$.}
\end{figure*}
  as even  flows that  in  their Eulerian description are smooth
present  messy float  trajectories \cite{bowen}.  
Typical ocean structures are eddies and currents. Major currents, such as 
the Gulf stream or the Kuroshio,  impact the Earth's climate because of the heat they transfer.  
 Eddies or rings  are certain robust, long-lived structures that may travel  hundreds to thousand of kilometres, and persist for periods lasting from months to years. 
 Understanding transport   across these large scale structures
is a challenging  task, but one increasingly amenable to treatment 
since  now data are becoming available  \cite{bowen}.
Lagrangian tools provide a characterization
of fluid flows. Underlying their description is Poincar\'e's idea of seeking geometrical structures 
on the ocean surface (the phase portrait) 
that can be used to organize particles schematically by regions corresponding to qualitatively different types of trajectories. 
For stationary flows the
 fixed points are  key for describing  the solutions geometrically. Fixed points
may be classified as hyperbolic or non-hyperbolic depending on their stability
properties. Hyperbolic fixed points are responsible for particle dispersion and 
non-hyperbolic fixed points are related to particle confinement. The interplay between 
dispersion and confinement  is an essential element of fluid transport processes. 
Stable and unstable manifolds of hyperbolic fixed points 
divide the phase portraits in regions with qualitatively different types of trajectories since they 
are barriers to transport. 
This letter  describes a new Lagrangian descriptor that for flows with a general time dependence realizes Poincare's idea of 
 dividing a  phase portrait in different regions that correspond to trajectories with qualitatively different 
behaviours. Our new instrument is based on a function which has been introduced in \cite{chaos} as a building block 
of a new definition of a  {\it Distinguished Trajectory} (DT), which is a generalization of the concept of fixed point
for aperiodically time dependent flows. 
Our function reflects, at the level of the  phase portrait, relevant dynamical features  of 
{\em arbitrary time dependent dynamical systems}.  Some of these features have not previously been detectable, 
thus,  when applied to altimetric ocean data sets, $M$ reveals the hidden geometry of the ocean flow. 
The technique
  locates simultaneously hyperbolic and non-hyperbolic flow regions. Since it 
  reveals singular features along  the  stable and unstable manifolds of the {\it Distinguished Hyperbolic Trajectories} (DHTs), it is
  useful as well to detect  these invariant curves, that near the DHTs coincide with its stable and unstable directions.  

{\it The function $M$}.---  The function we propose   as a global Lagrangian descriptor 
(see \cite{chaos}),  considers the system:
\begin{eqnarray}
\dot{{\bf x}} &=& {\bf v}({\bf x},t), \label{adv1}  \, \,{\bf x} \in \, \mathbb{R}^n, \, t \in \, \mathbb{R}
\end{eqnarray}
where  ${\bf v}({\bf x},t)$ is  $C^r$ ($r \geq 1$) in ${\bf x}$ and continuous in $t$.
Let ${\bf x}(t)$ denote a trajectory and 
denote its components in $\mathbb{R}^n$
by $(x_1,x_2,...,x_n)$.
For all initial conditions ${\bf x^*}$   in an open set ${\mathcal B}\in\mathbb{R}^n$, at a given time    ${ t^*}$,
we define the  function $M({\bf x^*}, t^*)_{{\bf v},\tau}:({\mathcal B}, t) \to\mathbb{R}$  for the system~(\ref{adv1}) as follows: 
\begin{equation}
M({\bf x^*}, t^*)_{ {\bf v}, \tau}=   \int^{t^*+\tau}_{t^*-\tau} \! \!\! \left( \sum_{i=1}^n \left({d x_i(t)}/{dt}\right)^2 \right)^{1/2} dt  \label{def:M}
\end{equation}
In our observational oceanographic flow, particle advection occurs 
mainly in  2D (see \cite{nlpg}), so $n=2$ in~(\ref{adv1}).
$M$ is then  the function that measures the Euclidean arc-length of the curve traced by a trajectory passing through 
 ${\bf x^*}$ at time  ${ t^*}$ on the plane $(x_1,x_2)$. The trajectory is integrated from ${ t^*}-\tau$ to ${t^*}+\tau$.
  The function $M$ depends  on  $\tau$ and also on the vector field {\bf v}.  It is applicable to both time dependent and stationary flows.
 In the latter case it provides a {\it time independent} partition of the phase portrait. For instance, for  the unforced, undamped Duffing equation
 a contour plot of  $M$  depicts the familiar stable and unstable 
manifolds of the fixed point located at the origin. For time dependent flows
 the phase space partition provided by $M$ is  {\it time dependent}.
  \begin{figure*}
\includegraphics[width=12cm]{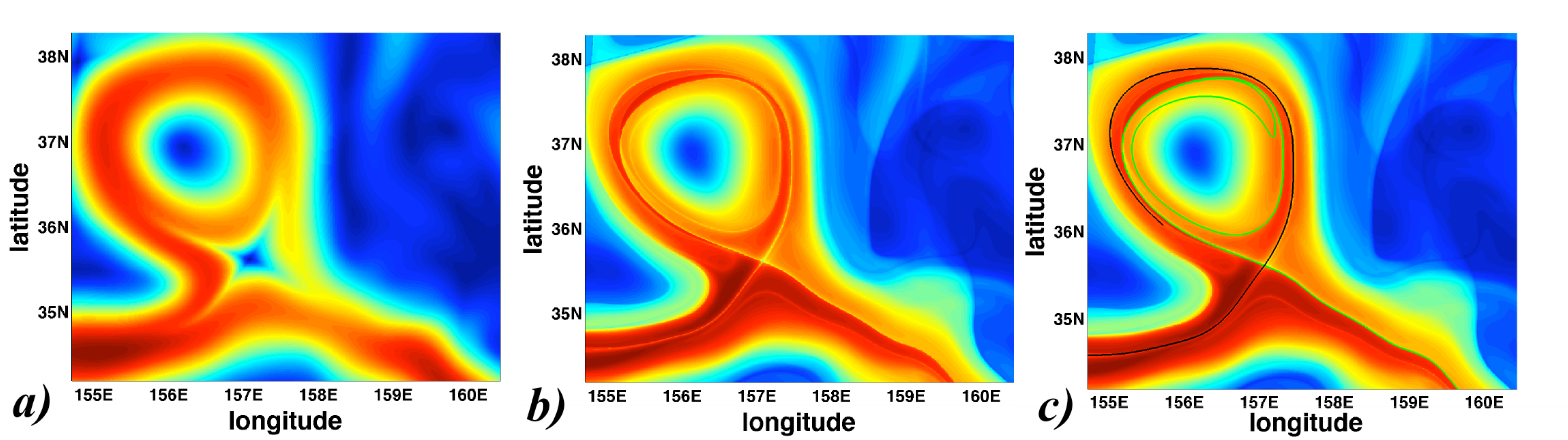}
\caption{\label{fig:M} The function $M$ on  May 2, 2003. a) $\tau=2$; b) $\tau=15$; c) the same as b) with a piece of stable manifold (black) and a piece of unstable manifold (green) of the DHT  overlapping.}
\end{figure*}

 {\it The data set and the dynamical system}.--- The velocity data used  in this 
work  are geostrophic
surface currents computed at  CLS Int Corp (www.cls.fr) in the framework of the
SURCOUF project \cite{larnicol}. 
The data span the whole Earth, at a resolution of 1/3 degrees at Equator, but we focus our results over  a region through which passes  the Kuroshio current,  in selected days
of May and June 2003. Details on the data  may be found in \cite{nlpg2}. It is interpolated  following methods described in \citep{JPO,msw}, that
use bicubic interpolation in space and Lagrange polynomials in time. Our
 coordinate system ($ \phi,\mu$), is related to the longitude and latitude  ($ \phi,\lambda$) by means of a transformation $\mu=\mu(\lambda)$
(see details in \citep{JPO,nlpg2}). These variables are convenient for they  distribute the data on a uniform grid.
The equations of motion   for ($ \phi,\mu$) are,
\begin{eqnarray}
\frac{d \phi}{d t}=\frac{u(\phi,\mu,t)}  {R \, {\rm cos}(\lambda(\mu))}, &&
\frac{d \mu}{d t}=\frac{v(\phi,\mu,t)}{R \,{\rm cos}(\lambda(\mu))} \label{sd2}
\end{eqnarray}
where $u$ and $v$ represent the eastward and northward
components of the altimetry surface velocity field { respectively},   and $R$ is the radius
of the Earth.  The factor,  $1/{\rm cos}(\lambda(\mu))$,  in the  $\mu$ equation is an artefact of the coordinate transformation.
The function $M$ in Eq.  (\ref{def:M}) is 
computed over the  dynamical system (\ref{sd2}). Thus 
the length of the trajectory is  measured on the  ($\phi, \mu$) plane. 
The system expressed in Eq. (\ref{sd2}) is not exact,  as it is subject to errors coming from the measured velocity fields,  the sort of interpolation used, etc. However Eq. (\ref{sd2})
is used for evaluating  $M$, a function that contains Lagrangian information. In the literature  \cite{hallerphf}, has been studied
 the robustness of the Lagrangian structures under errors induced in the vector field satisfying certain conditions. We have assessed the reliability of
 $M$  by computing it with several interpolation schemes.
 
  \begin{figure*}
\includegraphics[width=11cm]{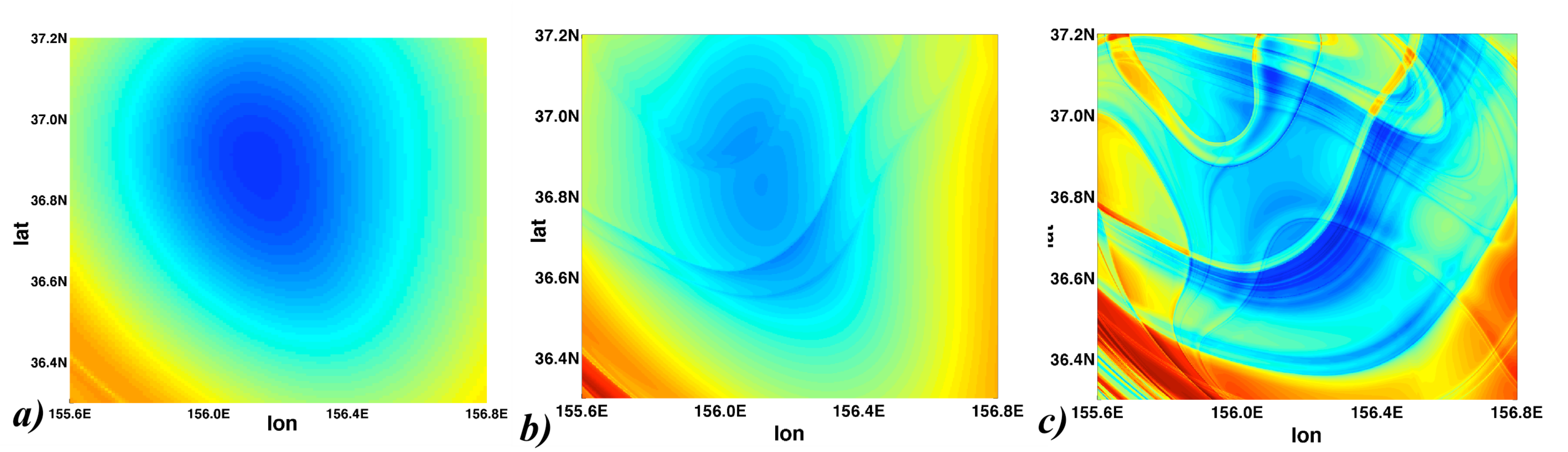}
\caption{\label{meddie} The function $M$ evaluated over the inner core of an eddy on May 2, 2003. a) $\tau=15$; b) $\tau=30$; c) $\tau=72$.}
\end{figure*}

 {\it Results}.--- We demonstrate that the function defined in
 Eq.  (\ref{def:M}) gives a global dynamic picture of  oceanic flows  since
 it   detects simultaneously  invariant manifolds, hyperbolic, and non-hyperbolic flow regions.
It synthesizes information more efficiently than for instance spaghetti diagrams. These represent paths over time of messy trajectories but they do not
communicate information about regions in which particle evolutions are qualitatively different, and one cannot get much intuition from
them.
 Figure \ref{fig:M1} displays the function $M$ for medium and large $\tau$ on selected days of May and June 2003 along the meandering Kuroshio current. 
 Maximum values of $M$ are in red, while dark blue indicates minima.
 The dependence of $M$ on time is obvious for this highly aperiodic flow,  since representations for different days have
different structures.  In the figure the organizing centres  are visible at a glance. These key points are the minima of $M$, and
 as discussed in \citep{chaos} they are related
 either to hyperbolic or non-hyperbolic distinguished trajectories. 
 Singular features of $M$ forming lines 
 are easily discerned, both in Figs. \ref{fig:M1} and  \ref{fig:M}. Fig. \ref{fig:M}b) shows their intersection
at a hyperbolic minimum at longitude $\sim 157.1^o$ and latitude $\sim 35.63^o$. 
 The time  evolution of this point has been characterized   in \cite{nlpg2} as a 
DHT.  Singular lines are identified as manifolds since they are advected by the flow and are asymptotically
 obtained from small segments aligned with the stable and unstable subspaces of the DHT. Fig. \ref{fig:M}c) shows the overlapping of $M$
with the stable and unstable manifolds  computed with the technique used in \cite{nlpg}.  This  confirms the coincidence of
the lines with the manifolds. Why should stable and unstable manifolds be traced out  by singular features of $M$? $M$ measures the lengths of  curves 
traced by trajectories on the phase space,  so it is expected it will change abruptly  at the boundaries of  regions  comprising trajectories with  qualitatively different evolutions, since this
is exactly what the stable and unstable manifolds
separate. Convergence of the structure of $M$ towards these singular lines requires a large enough $\tau$ value.
For instance for $\tau=2$ the appearance of  $M$ in Fig. \ref{fig:M}a) is rather simple, almost without structure and resembling that of Eulerian currents, while 
sharp lines in   Fig. \ref{fig:M}b) and c) require the use of  $\tau=15$.
The structure of $M$ becomes more and more refined
for larger $\tau$ values as confirmed by  panels  \ref{fig:M1}b) and d), obtained for $\tau=30$. This is justified  because $M$ reflects the history of initial conditions on open sets,
and in highly chaotic systems this history is expected to be more complex  for longer time intervals. 
The evaluation of $M$ in large oceanic areas as shown in Fig. \ref{fig:M1} reveals recognizable phase portraits 
similar to those of the cat's eyes  of the forced pendulum (in panel a) upper left), or 
the forced Duffing equation (see panel c) at the lower right). The ocean surface resembles a patchwork of interconnected dynamical 
systems from which the complexity of possible particle routes is visible. 
  \begin{figure*}
\includegraphics[width=13.cm]{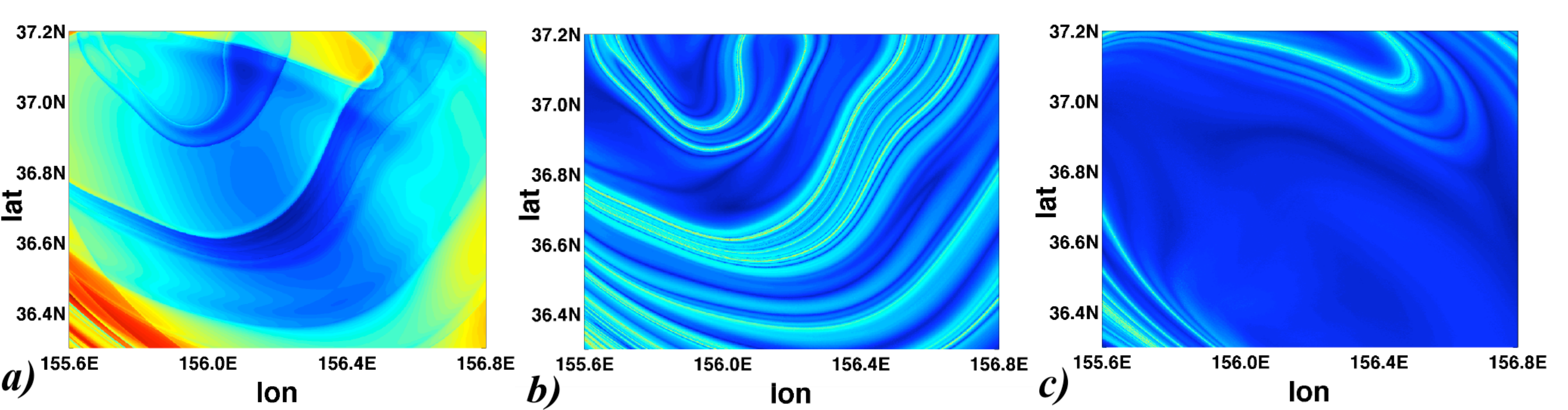}
\caption{\label{compare} Comparison of diverse Lagrangian techniques for $\tau=50$ at the same area of Fig. 3 ; a) The function $M$; b) the 
forward FTLE field; c) the  
backward FTLE field.}
\end{figure*}

In figures \ref{fig:M1} and \ref{fig:M}, apart from the minima of the function $M$
at the intersections  of singular lines, related to the hyperbolic DT,  there are apparent
minima at the eddy centres.  In the work by Madrid and Mancho \citep{chaos} 
these have been related to non-hyperbolic DT (DET), which
are  eddy-like structures, of great interest to oceanographers.
 The Lagrangian description of eddies, such as that shown in  Fig. \ref{fig:M}
  reveals the existence of an inner core, which is robust and rather impermeable to stirring and an outer ring, where the interchange with the media is understood in terms of lobe dynamics
(see \cite{physrep}).  
We analyze how the function $M$ reflects to what extent the inner core of Fig. \ref{fig:M} 
is impermeable to mixing. Figure  \ref{meddie} displays contour plots of $M$ on $t=$May 2, 2003
for several $\tau$. In Fig. \ref{meddie}a) it is observed that for  
$\tau=15$ days the interior of the eddy has a minimum   which is locally smooth. 
This means that  in the range $(t-\tau, t+\tau)$
trajectories in this neighbourhood outline similar paths,  and for this reason the function $M$
does not change sharply (i.e does not have singular features).
Smoothness of $M$  implies that  for these initial conditions  
it does not perceive  nearby hyperbolic regions for  $(t-\tau, t+\tau)$. 
Hyperbolic trajectories are the ones  responsible for dispersion and it is
just these trajectories  that may induce 
sharp changes in $M$.  In Figures  \ref{meddie}b) and c) for larger $\tau$ values (i. e, $\tau=30$ 
and 72 respectively) the interior of the eddy  becomes less and less smooth, for 
 in the range $(t-\tau, t+\tau)$ trajectories placed at the interior 
core  either were dispersed in the past or will disperse in the future. In fact, in Figure   \ref{meddie}c), the interior of the core 
is completely foliated by singular features associated either to stable or unstable 
manifolds of nearby hyperbolic trajectories.
So, the value at which  $M$ starts losing  smoothness,
e.g. $2 \tau=60$, is a good indicator  of the maximum time for confinement
 of particles in the inner core.
The minimum of $M$ on the elliptic region does not converge with $\tau$, and this is 
 the condition required for finding DT. Similarly to what is described in \cite{chaos}, 
 DET have not been found in highly aperiodic flows. 
 Figure  \ref{meddie}c)  displays  in  black line with a computed unstable manifold which overlaps on the contour plot of $M$.
 Again there is observed a coincidence of the singular features of $M$ with the manifold. However
 the foliated structure of $M$ is much  richer than that provided by the manifold. The reason is that the manifold has been computed from  
 the one  DHT  recognizable in Fig. 2b) and c),  while  $M$ displays all stable and unstable manifolds from all  possible DHTs in the neighborhood of the eddy,
 without need for identifying DHTs a priori, as required by the manifold algorithm (see   \cite{nlpg}). Thus $M$ provides a complete partition of the phase portrait, 
 while the  direct computation of a manifold of a DHT does not.

 The Lagrangian method using  $M$ has several advantages over other methods based on finite time versions of 
 Lyapunov exponents (LE) such as FTLE or FSLE \cite{hallerch,emilio}.  LE techniques  provide information
on the linearised  flow along trajectories and their focus is on hyperbolic regions. Ridges of 
FTLE and FSLE fields represent manifolds as reported for instance in \cite{hallerch,emilio}. 
Figure   \ref{compare} confirms this point. In it there is displayed   the same eddy of  Fig.  \ref{meddie}, but for $\tau=50$.
The sharpest ridges of the FTLE field represented in Fig. \ref{compare}b) and c) are
associated to the stable (forward FTLE) and unstable (backward FTLE) manifolds respectively,  in close correspondence to the singular features of
$M$ displayed in Fig. \ref{compare}a). An obvious difference between these two representations is that $M$ contains the information on both stable and unstable
manifolds in the same picture, while FTLE splits it in two diagrams. Regarding other features provided by  FTLE and $M$
there is not  strict agreement. In the regions centered at longitude $\sim 156.1^o$ and latitude $\sim 36.8^o$
and  at longitude $\sim 156.3^o$ and latitude $\sim 36.5^o$, the structure  of $M$ is smooth and eddy-like, which as explained above, indicates that particles in that area do not disperse.
The same areas in  Fig.  \ref{compare} b) display a striped pattern
suggesting that particles disperse in this time interval. Numerical integrations of trajectories in these regions confirm that they
stay close to each other in the interval  $(t-\tau, t+\tau)$. Non-sharp striped structures  in  Fig.  \ref{compare} b) raise then the questions of the kind of information they  provide and  if there is
an upper bound on $\tau $ for this vector field,  beyond  which  the validity of some of the structures provided by the FTLE cannot be  confirmed.  In \cite{brawigg} it is reported 
that for particular transient flows FTLE may develop `ghost' structures, although  a detailed  discussion on this is beyond the scope of our letter.
A further difference is that FTLE require certain  assumptions on the vector field (see \cite{hallerch}), while  by construction the function $M$ is defined for a general time dependent vector field
and there is required only  existence and  uniqueness of solutions for the system (\ref{sd2}).

Computationally the evaluation of the function $M$ is cheaper than that of LE.  In geophysical flows, both $M$ and LE  
require the performance of a large number of integrations on a dynamical system such as (\ref{sd2}), where the vector fields are interpolated over
a  finite space-time grid. Interpolations make computations expensive and saving these at each time step is a convenient feature. 
Evaluation of the function $M$ fits this criterion better than LE as  each point in the phase space  requires of just one integration forwards and backwards in time.
LE techniques however require more interpolations of the velocity field at each time step, either because they evaluate a separation rate among several trajectories, or
because they compute  the linearised flow around each trajectory.

The Lagrangian descriptor $M$  locates special organizing trajectories called DT as reported in \cite{chaos}. However although there are references
 suggesting the ability of LE to achieve this goal \cite{bramalek},  there are no published studies where this is discussed in detail.

In conclusion, this work  demonstrates the efficiency of a new Lagrangian descriptor $M$, for  
 identifying the essential dynamical features of general time dependent  flows. $M$ is a promising tool 
 for the estimation of transport in realistic flows, as recent articles
 \cite{nlpg2,jfm} have confirmed. Pursuing further the discussion on how it contributes to transport diagnosis, 
is beyond the scope of this letter. 


We thank D. Fox for useful comments, computer support from CESGA and  grants: Oceantech, 
I-Math C3-0104, MTM2008-03754, MTM2008-03840-E,  Simumat.


\bibliography{text_may}

\end{document}